\documentstyle[prb,aps,twocolumn,epsf]{revtex} 
\begin{document}
\title{Enhanced electrical resistivity before N\'eel order in the metals, RCuAs$_2$  (R= Sm, Gd, Tb and Dy)}
\author{E.V. Sampathkumaran,$^*$ Kausik Sengupta, S. Rayaprol and Kartik K. Iyer}
\address{Tata Institute of Fundamental Research, Homi Bhabha Road, Colaba,
Mumbai - 400005, India}
\author{Th. Doert and J.P.F. Jemetio}
\address{Technische Universität Dresden, Institut f\"ur Anorganische Chemie, Mommsenstrasse 13, D-01062 Dresden, Germany.}
\maketitle

\begin{abstract}
We report an  unusual temperature (T) dependent electrical resistivity($\rho$) behavior in  a class of ternary intermetallic compounds of the type RCuAs$_2$  (R= Rare-earths).  For some rare-earths (Sm, Gd, Tb and Dy) with negligible 4f-hybridization, there is  a pronounced minimum  in $\rho$(T)   far above respective N\'eel temperatures (T$_N$). However,  for the rare-earths which are more prone to exhibit such a $\rho$(T) minimum due to 4f-covalent mixing and the Kondo effect,  this minimum is depressed.  These findings, difficult to explain within the hither-to-known concepts, present an interesting scenario in magnetism. 
\end{abstract}
*E-mail address: sampath@tifr.res.in  
\vskip1.5cm

\maketitle

It is a fundamental fact in condensed matter physics that, in metals containing magnetic moments, the spin-disorder contribution to electrical resistivity ($\rho$) in the paramagnetic state is a constant adding to the lattice contribution, which decreases with decreasing temperature (T). Naturally, $\rho$ usually exhibits a positive T-coefficient down to the magnetic ordering temperature (T$_o$), except perhaps a weak critical point effect restricted to a narrow T-range (less than few percent of Curie temperature).\cite{1}  However, in the event that the sign of the exchange interaction between the local magnetic moment and the conduction electrons  is negative (the phenomenon called "the Kondo effect") due to covalent mixing of the relevant orbital, $\rho$ can show a low T-upturn increasing logarithmically below a characteristic temperature leading to a  minimum in $\rho$(T) well above T$_o$. This aspect has been very well known for Ce and Yb alloys.\cite{2} Here, we report that, in a class of ternary intermetallic compounds of the type RCuAs$_2$ (R= rare-earths),\cite{3} the rare-earths with strictly localized 4f character (in which case one does not anticipate the Kondo effect, e.g., Sm, Gd, Tb and Dy) exhibit a pronounced minimum in $\rho$(T)   above respective N\'eel temperatures (T$_N$).   However, for R= Pr, Nd and even Yb, no such minimum is observed in $\rho$(T). These findings present an unusual and interesting scenario as far as electronic transport in magnetic metals is concerned.  

The series of compounds under investigation, RCuAs$_2$ (R= Pr, Nd, Sm, Gd, Tb, Dy, Yb, and Y) crystallize in the HfCuSi$_2$-type layered tetragonal form (space group, P4/nmm)\cite{3} and the polycrystalline  materials in the single phase form were synthesized as described in Ref. 3.  The $\rho$ measurements (1.8 - 300 K) in zero magnetic field (H) as well as in the presence of a H of 50 kOe were performed on rectangular-shaped specimens (9mm X 2mm X 2mm) by a four-probe method employing spring-loaded pressure contacts with a typical spacing of 2mm between voltage leads. The sharp tip (diameter $<$0.2mm)  attached to the springs enables us to minimise the errors in $\rho$ arising due to uncertainties in the measurement of spacing between voltage leads.  In order to arrive at the T$_N$ values, we have also carried out dc magnetization (M)  (employing a commercial magnetometer) and heat-capacity (C) measurements.

Figure 1 shows    $\rho$(T) behavior for all the compounds below 70 K. Let us first look at the data for R= Sm, Gd, Tb and Dy, in which the 4f orbitals are known to be so deeply localized that the covalent mixing effects can be ignored. The value of $\rho$ falls in the milli-ohms range around 300 K in all cases. However, since the samples are found to be porous, the measured absolute values of $\rho$  are overestimated and hence should not be taken too seriously. What is important to note is that $\rho$ decreases monotonically with T   without any other feature down to 70 K (and hence not shown here). This establishes that all these compounds behave like metals. As the T is lowered, one finds that, in these compounds, $\rho$ attains surprisingly a prominent minimum at a temperature, T$_{min}$ (= about 35 K for Sm, Gd and Tb; about 20 K for Dy), below which it rises again. It should be noted that this feature is not observed for the Y compound (as well as for R= Lu, not shown here), in which case $\rho$(T) exhibits normally expected metallic behaviour down to the measured lowest  temperature (apart from a weak drop at about 7 K, which is presumably due to traces of a parasitic superconducting phase). This observation indicates that the minimum in other cases is magnetic in origin.

In order to rule out the origin of this upturn  from magnetic superzone formation\cite{4} due to possible antiferromagnetic ordering at T$_{min}$, it is important to make sure from independent measurements that T$_N$ falls well below T$_{min}$. In this respect, the dc magnetic susceptibility ($\chi$) and C data shown in Figs. 2 and 3 are quite conclusive. The $\chi$ above about 100 K is found to follow Curie-Weiss behavior (not shown here) and the effective moments ($\mu$$_{eff}$) obtained from this temperature range 
\begin{figure}
\centerline{\epsfxsize=7cm{\epsffile{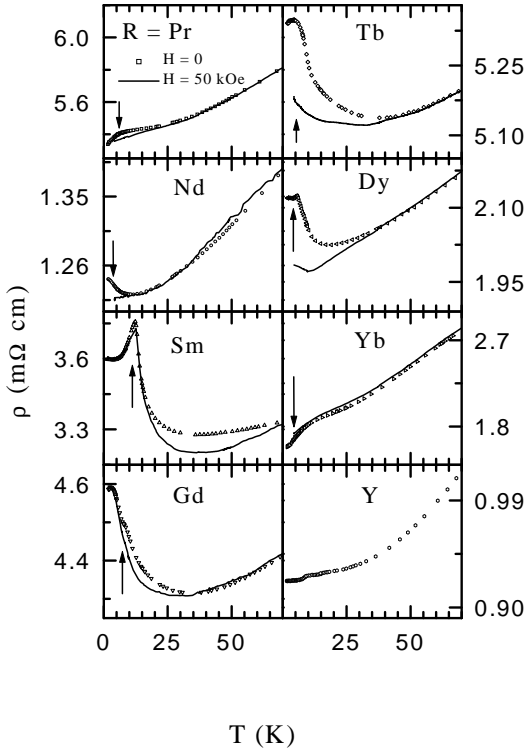}}}
\vskip5mm
\caption{Electrical resistivity behaviour of RCuAs$_2$ compounds below 70 K in zero magnetic field as well as in the presence of 50 kOe. Vertical arrows mark the N\'eel temperatures.}
\end{figure}
are in full agreement with that expected for respective trivalent R ions ($\mu$$_{eff}$) =  0.7, 7.9, 9.7, and 10.63 $\mu$$_{B}$)  for R= Sm, Gd, Tb and Dy respectively).    The features due to magnetic ordering appear as a well-defined peak in $\chi$ and C and/or as a sudden change in the slope of inverse $\chi$ versus T well below T$_{min}$ as one lowers the T.  The T$_N$ thus obtained are close to 12.5, 9, 9 and 8 K for Sm, Gd, Tb and Dy cases respectively. (There are additional features at further lower temperatures, which could be due to spin-reorientation effects. As this is not pertinent to the main conclusion of this article, we will not discuss this aspect further). There could be a small ambiguity in precisely locating T$_N$ from the C data, as the peak-positions can be shifted to a lower T or the C-tail may continue to a higher T, depending on the complexities of the magnetic structure.\cite{5} But the fact remains that the features due to the onset of magnetic transition appear well below T$_{min}$ in all cases.  Therefore, the minimum in   $\rho$(T) well above 12 K  can not be attributed to magnetic superzone gap formation. ($\rho$ increases further below T$_N$ in some cases, rather than exhibiting a drop due to the loss of spin-disorder contribution, which may be attributed to this gap effect).  It may be stated that $\chi$ for both the field-cooled and the zero-cooled conditions of the specimens are essentially the same, which establishes that there is no spin-glass freezing\cite{6} in the T range of investigation.  Isothermal M in the magnetically ordered state, measured till H = 120 kOe (see Fig. 2, bottom, for instance for the data at 1.6 K),  varies sluggishly with H without any hysteresis; in fact for R= Sm and Gd,  M varies linearly with H, whereas Tb and Dy exhibit  metamagnetic-like features.  These establish that the magnetic ordering is indeed of an antiferromagnetic-type and not of a ferromagnetic-type.
\begin{figure}
\centerline{\epsfxsize=7cm{\epsffile{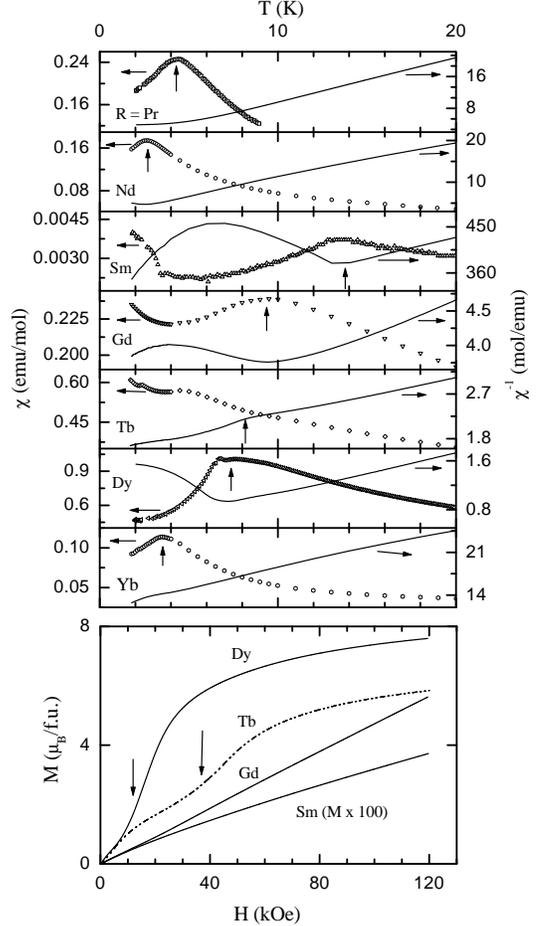}}}
\vskip5mm
\caption{(Top) Magnetic susceptibility measured in a magnetic field of 100 Oe (circles) and inverse susceptibility measured in a field of 5 kOe (lines) for RCuAs$_2$ compounds. Vertical arrows mark the N\'eel temperatures. (Bottom) Isothermal magnetization behavior for R= Sm, Gd, Tb and Dy at 1.6 K with the arrows showing the field around which there is an upward curvature in some cases and the plots are found to be non-hysteretic.}
\end{figure}
It may be recalled that we have earlier reported a similar $\rho$ behavior for some Gd systems, e.g., Gd$_2$PdSi$_3$, crystallizing in a AlB$_2$-derived ternary hexagonal structure,\cite{7} but other rare-earth members of those series do not exhibit such a minimum in   $\rho$(T). The present series of compounds is thus unique in the sense that the observation of $\rho$$_{min}$ spans over the entire rare-earth series (but not restricted to Gd alone). It is this finding that provides compelling evidence for a more-common hither-to-unrecognised magnetism-related electron scattering effects in the paramagnetic state. There is also another noteworthy difference between these two classes of compounds: The minimum is very sensitive to the presence of an external H in the case of Gd$_2$PdSi$_3$, whereas the application of a H of 50 kOe does not depress this minima in the present compounds, most notably for R= Gd and Sm (see Fig.  1). In other words, the magnetoresistance (MR= [$\rho$(H)- $\rho$(0)]/ $\rho$(0)) is quite small varying sluggishly with H; this also rules out any explanation in terms of possible grain boundary effects at least for R= Sm and Gd, as any extra scattering  from the magnetic ions at the grain boundaries have been in general known to be strongly suppressed by small applications of H. Thus, for H = 50 kOe, the magnitude of MR in the T-region of interest is much less than or close to 2$\%$, which could be compared with the net increase (about 14$\%$, 6$\%$, 4$\%$ and 4$\%$ for R= Sm, Gd, Tb and Dy respectively) of zero-field $\rho$ as measured at T$_N$ relative to that at T$_{min}$. Though smaller in magnitude, there is a noticeable influence of H for R= Tb and Dy. Possibly, there is a subtle difference in the origin of   $\rho$(T) minimum between the present Gd and Sm compounds on the one hand and  Gd$_2$PdSi$_3$ on the other with Tb and Dy compounds lying somewhere between these two limits.   
\vskip1cm
\begin{figure}
\centerline{\epsfxsize=6cm{\epsffile{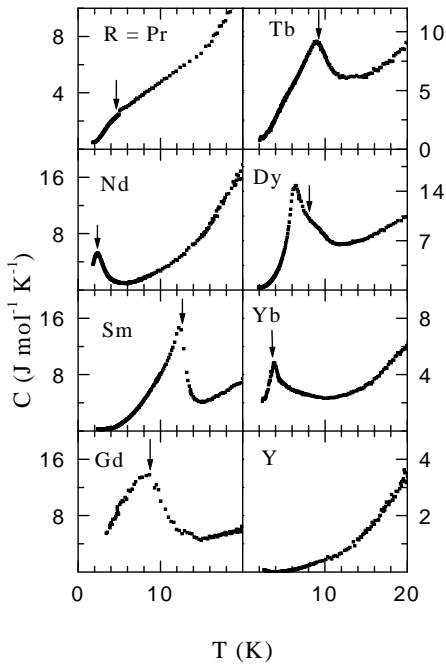}}}
\vskip1cm
\caption{Heat-capacity as a function of temperature for RCuAs$_2$ compounds. Vertical arrows mark the N\'eel temperatures, as inferred from the magnetic susceptibility data.} 
\end{figure}
In order to explore whether the upturn in $\rho$ below T$_{min}$ in the paramagnetic state could be understood in terms of hither-to-known concepts, we plot the data in the range 12.5 to 31 K (region of interest) in various ways in Fig. 4, for instance, for SmCuAs$_2$. Clearly, the upturn cannot be attributed to the Kondo effect due to deeply localized nature of the 4f orbital. Though, occasionally, Sm compounds show the Kondo effect,\cite{2} this phenomenon can not be expected for Gd, Tb and Dy.  The absence of the Kondo effect even in this Sm compound  is further supported by the non-logarithmic variation of $\rho$ (Fig. 4d). To substantiate this further, we have obtained the 4f contribution ($\rho$$_{4f}$) to $\rho$ by subtracting the lattice part from the knowledge of $\rho$(T) of the Y analogue. We would like to emphasize that the slopes of $\rho$(T) plots for these two compounds are found to be the same at room temperature, which implies that no further correction as described by Cattaneo and Wohlleben\cite{8} needs to be carried out to obtain precise lattice contribution.  The $\rho$$_{4f}$  thus obtained is nearly constant (see Fig. 4) at  temperatures above 100 K (consistent with the constancy  of spin-disorder contribution), however raising  gradually with  decreasing T in a non-logarithmic way. Since all the samples were synthesized under identical conditions and the pressure employed to obtain the compacted and sinteted pellets for $\rho$ studies in all samples are the same, we believe that possible errors in the $\rho$$_{4f}$(T) due to differences in porosity and consequent grain boundary effects among these samples are rather negligible. In addition, we arrive at a similar conclusion even when we employ the data of Lu sample for lattice contribution.  The plot of ln($\rho$) versus 1/T is not linear (Fig. 4c) and the absence of activated behavior rules out the formation of any type of gap as a possible cause of this upturn. The plots of ln($\rho$) versus T$^{-1/4}$ and T$^{-1/2}$ are also not linear (Figs. 4a and b), thereby ruling out an explanation in terms of variable range hopping and Coulomb gap formation mechanisms.\cite{9} Apparently, there is no T$^{1/2}$ dependence of $\rho$ as well (Fig. 4e), which implies that presently known  weak-disorder-induced electron localization ideas discussed by Lee and Ramakrishnan\cite{10} are not adequate to describe the anomalies. Another mechanism that could be advanced is in terms of the formation of magnetic polarons, as proposed for  EuSe (Ref. 11),  EuB$_6$ (Ref. 12) and Tl$_2$Mn$_2$O$_7$ (Refs. 13-15), but the application of a H is expected to have a pronounced depressing effect on the $\rho$-upturn resulting in a large negative MR, in contrast to the observation. For the same reason, the s-f exchange model (considering a strong scattering of 5d conduction electrons by the localized 4f electrons)\cite{16} proposed to explain the large MR behavior of GdI$_2$ may not be applicable; additionally, this model assumes ferromagnetic ordering at low temperatures, whereas the present compounds are antiferromagnets; it is not clear whether this model can be modified to explain the features. The fact that $\rho$ behavior in Y and Lu analogues is normal in this regard establishes that the origin of the above $\rho$ anomaly does not lie in the band structure.   Thus, there is no straightforward explanation for the above-mentioned  enhanced scattering, though it is clear that magnetism plays a role. 
\begin{figure}
\centerline{\epsfxsize=6cm{\epsffile{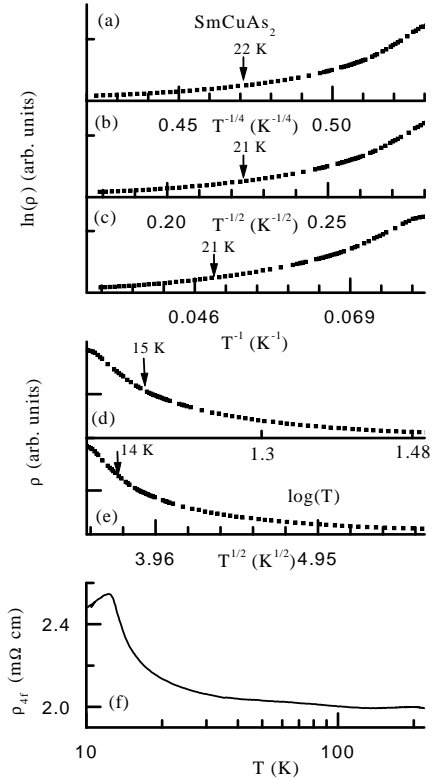}}}
\caption{The electrical resistivity in the temperature range 12.5-31 K plotted in various ways for SmCuAs$_2$ compounds. Fig. (f) shows the 4f contribution in the logarithmic temperature scale in the T range 10-225 K.}
\end{figure}  
We now inspect the $\rho$ behaviour of Pr, Nd and Yb compounds (Fig. 1) along with the $\chi$ and C behavior (Figs. 2 and 3). PrCuAs$_2$ exhibits a distinct feature due to antiferromagnetic ordering at about 4 K, as evidenced by the feature in $\chi$ (though the corresponding anomaly in C is not sharp presumably due to low entropy of the crystal-field-split ground state). Nd and Yb compounds are found to order magnetically close to 2.5 and 4.0 K respectively as marked by the features in C and $\rho$ data. It is worth mentioning that, though the T$_N$ values in this series are in general larger than  that expected on the basis of de Gennes scaling, this Yb compound turns out to be one of the few Yb compounds exhibiting  a high magnetic ordering temperature.\cite{17}  This makes this Yb compound interesting in its own right.  Following the behaviour in other members of this series described above, one would naively expect that there should be a minimum in the plot of  $\rho$(T) above T$_N$. However, it is striking to note that no such minimum could be observed for Pr and Yb compounds and the MR is also negligible; in other words, these compounds behave in a normal way in this respect. For the Nd compound, there is a weak upturn below 15 K to the tune of 1$\%$ only. We speculate that, at least for these cases, the origin of this suppression of $\rho$$_{min}$ may lie in non-negligible degree of 4f-mixing, characteristic not only of Yb, but also of Pr and Nd under favorable circumstances,\cite{18,19} though there could be other controlling factors.\cite{20} In the case of Yb compound, this covalent mixing manifests itself usually as the Kondo effect; the observed behavior of $\rho$(T) plot above T$_N$ is typical of Yb-based Kondo lattices\cite{21} and the temperature (around 130 K) at which $\rho$ starts falling as T is lowered represents the onset of coherent scattering among the Kondo centres.\cite{22} The positive sign of MR (see Fig. 1) arising out of destruction of Kondo-coherence by the application of H is consistent with this explanation.\cite{22} Therefore, we tend to believe that the Kondo-coherence somehow suppresses the upturn in $\rho$ as T$_N$ is approached. In the case of Pr and Nd compounds, the spatial mixing of 4f  (see Cooper\cite{2} and Ref. 19), though does not result in the Kondo effect, is apparently effective in depressing $\rho$$_{min}$. It is quite well-known\cite{18} that the strength of 4f-hybridization decreases progressively as one moves from Ce to Sm and the observation of a similar trend\cite{23} in $\rho$$_{min}$ supports this interpretation, atleast for these cases. 

To conclude, we report a new class of rare-earth based intermetallic compounds, RCuAs$_2$ (R= Sm, Gd, Tb and Dy), with a novel electrical transport behaviour before antiferromagnetic order.  The findings imply that the understanding of the electronic transport in the paramagnetic state of relatively simple magnetic metals is far from complete. We propose that the ideas based on weak localisation\cite{10} may have to be extended {\it incorporating strongly enhanced  magnetic critical scattering effects extending over a wide temperature range above T$_o$}.



\begin{references}

\bibitem{1}S.K. Ma, Modern.  Theory of Critical Phenomena, Benjamin, London, 1976; M.E. Fischer,   and J.S. Langer,   Phys. Rev. Lett. {\bf20}, 665 (1968).

\bibitem{2}B.R. Cooper,   Proc. High-Tc superconductors: Magnetic interactions, eds. L.H. Bennett et al (Singapore: World Scientific), 7 (1989); see also, for a review, M.B. Maple et al,  Handbook on the Physics and Chemistry of Rare Earths, eds. K.A.  Gschneidner, and L.  Eyring,  North-Holland, Amsterdam, 1978, p 797. 

\bibitem{3}J.-P. Jemetio et al   J. Alloys and Compounds {\bf338}, 93 (2002); J.-P.F. Jemetio et al, Z. Kristallogr. NCS, 
{\bf217}, 445 (2002). 

\bibitem{4}S. Legwold,  Magnetic properties of Rare-earth Metals, ed. Elliott  (Plenum, 1972).

\bibitem{5}J. A. Blanco et al, Phys. Rev. B {\bf43}, 13145 (1991). M. Bouvier et al, Phys. Rev. B {\bf43}, 13137(1991).  

\bibitem{6}K. Binder and A.P.  Young,  Rev. Mod. Phys. {\bf58}, 801-976 (1986). 

\bibitem{7}R. Mallik et al Eur. Phys. Lett., {\bf41}, 315 (1998); E.V. Sampathkumaran and R. Mallik, arXiv:Cond-mat/0211625 (to be published in "Concepts in Correlations" eds. A. Hewson and V. Zlatic (Klower Academic Publishers, The Netherlands, 2003).  

\bibitem{8}E. Cattaneo and D. Wohlleben, J. Magn. Magn. Mater. {\bf24}, 197 (1981). 

\bibitem{9}N.F. Mott,  Metal-Insulator transitions.  Taylor and Francis, London, 1990, 1

\bibitem{10}P.A. Lee and T.V. Ramakrishnan, Rev. Mod. Phys. {\bf57}, 287 (1985). 

\bibitem{11}Y. Shapiro et al,  Phys. Rev. B {\bf10}, 4765  (1974).

\bibitem{12}P. Nyhus et al, Phys.     Rev. B {\bf56}, 2717 (1997). 

\bibitem{13}Y. Shimakawa et al,  Nature (London) {\bf379}, 53 (1996).

\bibitem{14}M.A. Subramanium et al,  Science {\bf273}, 81 (1996). 

\bibitem{15}P. Majumdar and P.B.  Littlewood, Nature {\bf395}, 479 (1998). 

\bibitem{16}I. Eremin et al, Phys. Rev. B {\bf64},  064425 (2001).

\bibitem{17}See, for instance, D. Kaczorowski et al Phys. Rev B {\bf60}, 422 (1999); O. Trovarelli et al  Phys. Rev. B {\bf60}, 1136 (1999) 

\bibitem{18}R.D. Parks et al, Phys. Rev. Lett. {\bf52}, 2176 (1984); E.V. Sampathkumaran et al  Solid State Commun. {\bf55}, 977 (1985); Yu. Kucherenko et al  Phys. Rev. B {\bf65}, 165119 (2002) and references therein.

\bibitem{19}E.V. Sampathkumaran and I. Das,  J. Phys.: Condens. Matter. {\bf4}, L475 (1992)

\bibitem{20}R. Mallik and E.V. Sampathkumaran, Phys. Rev. B {\bf58}, 9178 (1998).  

\bibitem{21}See, for instance, E.V. Sampathkumaran et al   Physica B {\bf{186-188}}, 485 (1993).  

\bibitem{22}See, for  reviews, F. Steglich,  J. Magn. Magn. Mater. {\bf100} (1991); U. Rauschwalbe, Physica B {\bf147}, 1 (1987). 

\bibitem{23}Our investigations reveal that the Ce compound is dominated by the Kondo effect only in the range 2 - 300 K. 



\end{references}
\end{document}